\documentclass[ 
aps,%
11pt,%
final,%
notitlepage,%
oneside,%
onecolumn,%
nobibnotes,%
nofootinbib,% 
superscriptaddress,%
noshowpacs,%
centertags]%
{revtex4-2}

\usepackage[cp1251]{inputenc}
\usepackage[T2A]{fontenc}
\usepackage[dvips]{color}
\usepackage{amsmath,amsfonts,amssymb,amsbsy,epsf,colordvi,rotate,array}
\usepackage[mathscr]{eucal}
\usepackage{hyperref}
\usepackage{graphicx}

\begin{document}

\title{Reciprocity in Heavy Quark Fragmentation Function}
	
\begin{abstract}
At high energies a form of fragmentation function for a heavy quark into a hadron is substantiated to agree with a reciprocity relation to the distribution of heavy quark as the virtual parton in the hadron. The relevance of the relation is analysed in its application to empirical data.
\end{abstract}

	\author{\firstname{V.V.}~\surname{Kiselev}}
	\email{kiselev.vv@phystech.edu; Valery.Kiselev@ihep.ru}
	
	\affiliation{Moscow Institute of Physics and Technology,
		Institutsky 9, Dolgoprudny, Moscow Region, 141701, Russia}
	\affiliation{Institute for High Energy Physics named by A.A.~Logunov of National
Research Centre «Kurchatov Institute», Nauki 1, Protvino, Moscow Region, 142281, Russia}	
	
	\maketitle
\section{Introduction}	
A hadron composed of heavy quark and light degrees of freedom like light quarks and gluons can be considered in the fast velocity frame so that partons bound in the hadron are characterized by fractions of hadron momentum $x=p_\mathrm{part}/p_H$, while the binding energy is much less than the momentum of hadron $p_H$. In such a picture all of partons are moved with the same 4-velocity $u^\mu$ equal to the 4-velocity of the hadron. However, the light degrees of freedom form a lump with a spread  invariant mass, which is reflected in the heavy quark distribution.

We use this picture in Section \ref{sec-2} to describe the distribution of heavy quark $f_Q(x)$ in the hadron as well as a relevance to Kuti--Weisskpopf model \cite{KW}. In Section \ref{sec-3} we consider the fragmentation function of heavy quark into the hadron $f_H(x)$, wherein $x=p_H/p_Q$,  and substantiate the reciprocity relation $f_H(x)=f_Q(x)$ as well as we discuss on a region of its validity \cite{KLP}, which involves hard gluon corrections. In Section \ref{sec-4} a comparison with experimental data is presented. The results are summarized in Conclusion.

\section{Quark-gluon Lump and partons\label{sec-2}}
\begin{figure}[t]
\begin{center}
\includegraphics[width=4.in]{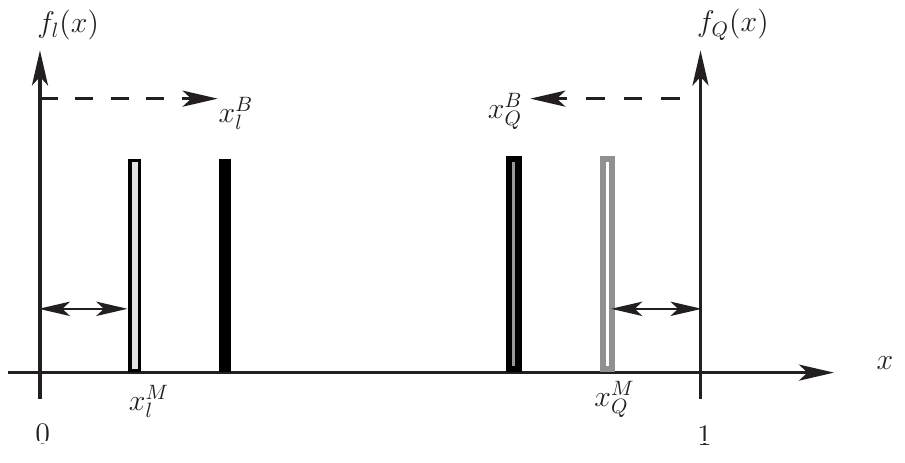}
\caption{Specular on mass-shell distributions of light  and heavy partons $f_l(x)$ and $f_Q(x)$, respectively, in the naive model of heavy meson mark by superscript $M$ and the heavy baryon marked by superscript $B$.}
\label{naiv}
\end{center}
\end{figure}

In the very simplest initial iteration the hadron in question is composed by two valence partons in a meson as the heavy quark and light antiquark or by three partons in a baryon as the heavy quark and light di-quark, so that all of such partons are on mass shells and possess a common 4-velocity $u^\mu$ equal to the velocity of the hadron $u^\mu_H$. Therefore, the distribution of heavy quark versus the fraction of hadron momentum is specular to the distribution of light component: the light component is posed at 
$$
	x_l=\frac{m_l}{m_H},
$$
while the heavy quark is posed at mirrored value
$$
	x_Q=\frac{m_Q}{m_H}=1-x_l.
$$
These distributions are schematically shown in Fig. \ref{naiv}. Since the constituent di-quark in the baryon is considered to be twice massive than the constituent quark in the meson, the  distribution of heavy quark in the heavy baryon is essentially softer than that is in the heavy meson. 

\begin{figure}[bh]
\begin{center}
\includegraphics[width=4.5in]{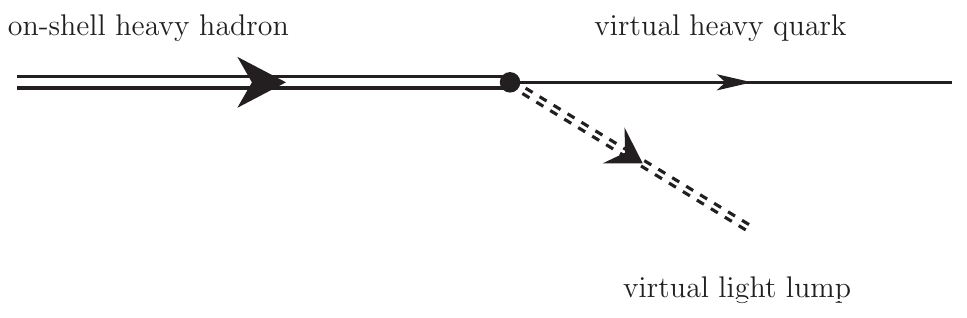}
\caption{Composing the heavy hadron by virtual light lump and virtual heavy quark.}
\label{virt1}
\end{center}
\end{figure}

In the heavy hadron being the bound state the light components and the heavy quark are not on mass-shell because of the strong interaction. The virtuality of light lump can be treated as the invariant mass $\mu_l(x)$ spread versus the fraction of heavy hadron momentum so that $0<\mu_l(x)<m_H$. Therefore, one can introduce the probability distribution of light lump versus the virtuality, that originates both the distribution of light lump as probability versus $x=\mu_l(x)/m_H$ and its specular probability for the virtual heavy quark $f_Q(x)\approx f_l(1-x)$, since the hadron is on mass-shell, hence, $\mu_l(x)+\mu_Q(x)=m_H$ by the very definition of virtual components of hadron in the model with identical 4-velocity of these components motion $u^\mu$ equal to the velocity of heavy hadron $u^\mu_H$. Therefore, we explore the law of 4-momentum conservation in the fast velocity frame: 
\begin{equation}
\label{4-law}
	p_M^\mu=\mu_Q u^\mu+\mu_l u^\mu\;\Rightarrow\;
	m_H u^\mu_H=\mu_Q u^\mu_H+\mu_l u^\mu_H\;\Rightarrow;
	m_H=\mu_Q+\mu_l\;\Rightarrow\; 1=x_Q+x_l.
\end{equation}
This statement is illustrated in Fig. \ref{virt1}.

Note that the model of equivalent rapid velocities follows from the soft non-perturbative picture in the rest frame of heavy meson: in kinematical $s$-channel the meson decays to virtual heavy quark with mass $\mu_Q$ and virtual light lump with mass $\mu_l$ at soft spatial momenta, so that at fast velocity frame one can neglect these soft momenta. In this pattern the probabilities for virtualities could be arbitrary, say, one could choze arbitrary distribution of probability versus $\mu_l$, in principle. However, we can consider the situation with the most probable distribution, so that the probability to find the on mass-shel heavy meson under actual distributions over the virtualities is equal to one. This approach fixes the condition of extremity for relevant distributions fixed for the maximal probability.

The distribution of light lump $f_l(x)$ is not the distribution of light partons. So, one studies the parton distributions, which results in different parton functions for see and valence quarks as well as for gluons. However, the heavy quark distribution obeys two certain limits for partons: at $x\to 0$ one reaches the Regge approximation in the scattering processes, while at $x\to 1$ one arrives to the deep elastic processes with rigid limits depending on the number of constituents in the hadron as was described by Kuti--Weisskopf model \cite{KW}. These limits establishes quantitive description for differences in distributions for mesons and baryons as well as a more hard or leading character of parton functions for heavy quarks \cite{Kartvel2,Chliapnikov,Kartvel} as the indicator of dynamical breaking of flavour symmetry in strong interactions. Thus, one find
\begin{equation}
\label{fQ}
	f_Q(x)\sim x^{-\alpha_Q}(1-x)^{\gamma_H},
\end{equation}
where $\alpha_Q$ is determined by the heavy quark intercept of Regge trajectories for heavy hadrons, and $\gamma_H$ correlates with the rule counting the number of hard gluons necessary for elastic scattering of composite hadron system. Value of $\gamma_H$ is greater for baryons in comparison to mesons, while $\alpha_Q$ is negative for charm and beauty quarks, and $|\alpha_b|>|\alpha_c|$. So, the most probable distribution of light lump rigorously correlates with the behaviour of heavy quark as the parton. 

The estimates of $\alpha_Q$ parameters can be obtained beyond the Regge treatment. So, in \cite{Kartvel3} authors considers heavy quarkonia $(Q\bar Q)$ as composed by the heavy quark and heavy anti-quark. Therefore, a distribution versus the heavy quark virtuality is determined by a non-relativistic wave function of the quarkonium. The wave function normalizes a leptonic decay width of the quarkonium, hence, one gets the phenomenological method to extract the wave function parametrization in terms of $\alpha_Q$ in the fast velocity frame like 
$$
	\psi_{Q\bar Q}(x)\sim x^{-\alpha_Q}(1-x)^{-\alpha_Q}.
$$
These estimates of $\alpha_Q$ are in admitable agreement with those obtained in the Regge trajectories approach. However, the question is whether the virtuality of heavy quark in the heavy quarkonium to be identical to the virtuality of heavy quark in the heavy-light hadron. Nevertheless, one could expect that the leading features of distributions versus the heavy quark fraction of hadron momentum are matched by such a growing $|\alpha_Q|$ parameter with the increase of heavy quark mass. 

Another issue is a fine dependence of virtualities on the sort of heavy hadron with the same quark content, say, the difference in distributions for vector and pseidoscalar heavy mesons appearing due to a spin-spin inteactions. We expect that the spin is more rigorous characteristic for the heavy quark that for the light lump, hence, more mass of vector meson $m_{H^*}$ means more virtual mass of heavy quark $\mu_{Q^*}$. So, we consider that the heavy quark distribution in the vector meson is slightly harder than in the pseudoscalar meson.

\section{Heavy quark fragmentation\label{sec-3}}
\begin{figure}[th]
\begin{center}
\includegraphics[width=4.5in]{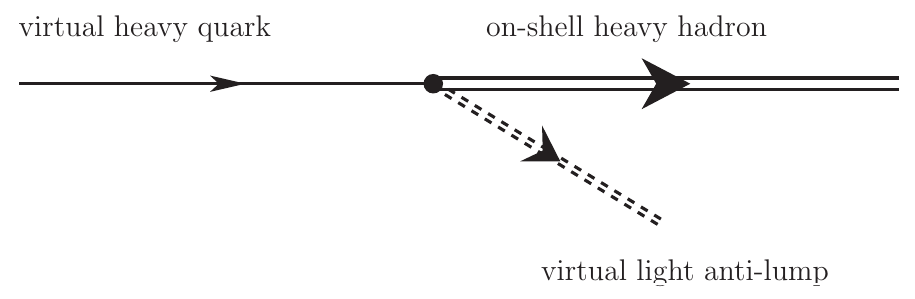}
\caption{Fragmenting the virtual heavy quark into the on mass-shell heavy hadron and  virtual light lump.}
\label{virt2}
\end{center}
\end{figure}

The fragmentation is considered in hard inclusive processes when fast virtual heavy quark transforms to the heavy hadron. Consider a soft stage of fragmentation, when the virtual heavy quark with invariant mass $\mu_Q^F$ in its rest frame non-perturbatively decays to the on mass-shell heavy hadron and a virtual light anti-lump with virtuality $\mu_{\bar l}$. Due to the softness the process in a fast velocity frame corresponds to the situation, when all participants possess the same 4-velocity $u^\mu$ equal to the 4-velocity of virtual heavy quark $u^\mu_Q$ (see Fig. \ref{virt2}). Therefore, the conservation of 4-momentum results in 
\begin{equation}
\label{4-law-2}
	\mu_Q^F=m_H+\mu_{\bar l}\;\Rightarrow\; 1=x_H+x_{\bar l},
\end{equation}
wherein the fractions of momenta are introduced in the standard manner: $x_H=m_H/\mu_Q^F$ and $x_{\bar l}=\mu_{\bar l}/\mu_Q^F$.

\begin{figure}[b]
\begin{center}
\includegraphics[width=4in]{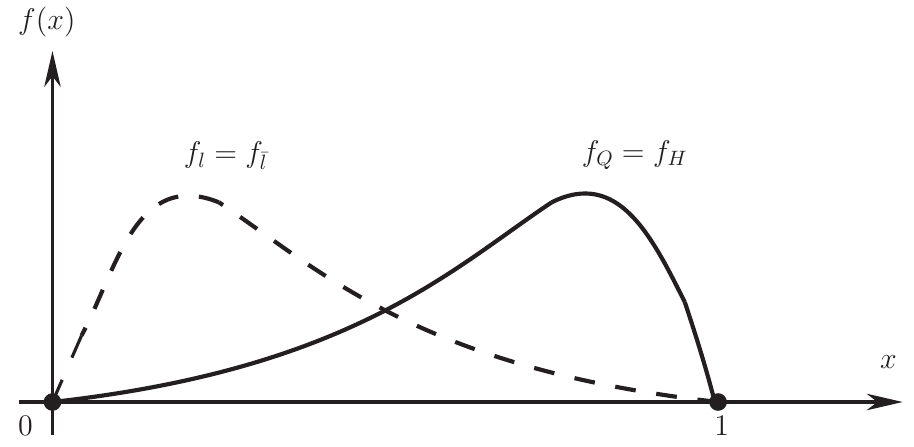}
\caption{The reciprocity as the specular relation for the light and heavy virualiries and momentum fractions.}
\label{recip}
\end{center}
\end{figure}

One can straightforwardly conclude that the probability distribution versus the vurtuality of light anti-lump specularly determines the distribution of heavy hadron,
\begin{equation}
\label{rec}
	f_H(x_H)\approx f_{\bar l}(1-x_H)
\end{equation}
This raises the question: what is the maximally probable form of $f_{\bar l}(x_{\bar l})$? In order to answer this question note that we have already known that the most probable distribution of light anti-lump relevant to the soft strong interactions with certain heavy hadron corresponds to it incoming into the heavy anti-hadron, because such the distribution gives the unit probability of the process\footnote{One can imagine that the light anti-lump composes the heavy anti-hadron due to fusion with an appropriate virtual heavy anti-quark. The probability of such soft fusion is equal to one if the distribution of light anti-lump versus the virtualities is optimal and it agrees with the maximal probability. In this pattern two virtual quark transfer to two heavy hadrons, hence, more masses heavy hadrons mean more virtualities of heavy quarks. Thus, the distribution of heavy quark in the vector meson, say, should be harder than one in the pseudoscalar meson.}. Other soft channels can not destroy the most probable distribution of virtualities for the light anti-lump, but the maximally probable composition of the light anti-lump determines the distributions of  light partons inside the lump that results in various variants of further soft hadronization of these partons. Thus,
\begin{equation}
\label{bar-l}
	f_{\bar l}(x)=f_l(x),
\end{equation}
hence, the distribution of heavy hadron in the soft stage of heavy quark fragmentation repeats the distribution of heavy quark in the heavy hadron (see Fig. \ref{recip}),
\begin{equation}
\label{frag1}
	f_H(x)=f_Q(x).
\end{equation}

The soft stage means that the virtuality scales are given by characteristic momentum transfers of heavy quark inside the heavy hadron, $\mu(H)\sim p_\perp(H)$. In hard processes at scales of $p_\perp\gg \mu(H)$ one has to account for the evolution of fragmentation function versus virtuality $\mu_\mathrm{hard}$ in terms of hard gluon and quark emission before the soft stage.

\section{Relevant empirics\label{sec-4}}
The definition of fast velocity frame for the fragmentation itself implies that the ratio of heavy hadron mass to the maximal kinematical energy of the hadron tends to zero, $x_\mathrm{min}=m_H/E_\mathrm{max}\to 0$. Therefore, data at a finite ratio $x_\mathrm{min}$ could be treated only approximately with a systematic uncertainty in the $x$ position given by $\Delta x\approx x_\mathrm{min}$.  Such an uncertainty admits to see only the effect of leading for the charm-quark in the fragmentation to charmed hadrons: the hadron acquires a hard fraction of heavy quark momentum with the most probability, indeed (see data by ARGUS \cite{ARGUS}, CLEO \cite{CLEO} and BELLE \cite{Belle}, when $\Delta x\sim 0.3$). Unfortunately, in this situation one has got no chances to consider the data quantitatively as concerning for any parameters like $\alpha_Q$. 

At high energies the physics of fragmentation is divided into two stages. The first stage is a hard production of heavy quark where one  has to account for a perturbative production at a leading order and next-to-leading orders in a strong coupling constant with an inclusive  emission of gluons and light quarks. The second stage is the soft fragmentation presented in the previous sections. The result of these stages is calculated as a convolution of core for the perturbative evolution from the hard scale to the soft one with the soft fragmentation function \cite{FieldRD}. The evolution equations for fragmentation function $D_Q^H(x,\mu)$ at a hard scale $\mu$ take the form 
$$
	\frac{\partial D_Q^H(x,\mu)}{\partial\ln \mu}=\sum_p\int\limits_x^1\frac{\mathrm d z}{z}\,
	\mathscr P_{p\to Q}(x/z,\mu)\,D_Q^H(z,\mu),
$$
where $p$ marks a paron splitting to the heavy quark $Q$ in the perturbative regime, $\mathscr P_{p\to Q}$ denotes a splitting kernel. The evolution makes the distribution to be more softer. 
Therefore, one has not got a direct information about the soft fragmentation, since it is screened by the necessary hard evolution required for the consideration in the fast velocity frame. Particularly, one can explore various  parameterisations, say, as a soft term by Peterson's fragmentation function in \cite{Peterson}.

In this context one can refer to results obtained in the theoretical calculations presented in \cite{Cacciari,Bonino,Cacciari1}, wherein both stages of heavy quark fragmentation are taken into account. 

The perturbative calculations are more justified for the fragmentation of heavy quark into the heavy quarkonium \cite{BBL}, which can be used as a laboratory of such processes since a production of companion heavy quark is a hard process itself. In this case the virtualities obey quite narrow region in comparison to the hard scale of quark production that can be fitted by non-relativistic wave function if the pair of heavy quark and heavy anti-quark is in singlet state with respect to gauge group of quantum chromodynamics (QCD), while some corrections appear due to suppressed coloured operators in the framework of effective theory such as non-relativistic QCD (NRQCD) \cite{NRQCD} or potential NRQCD (pNRQCD) \cite{Brambilla1,Brambilla}. The corrections are significant in a hadronic production of fragmenting heavy quark, when the while in colour and coloured operators contribute with different dependence versus inverse transverse momentum.

The fragmentation in electron--positron annihilation allows us to study the heavy quarkonium almost  perturbatively at large energies when one can neglect the corrections due to the coloured states of heavy quark and heavy anti-quark pair. The analityc calculations of heavy quarkonium fragmentation functions for vector and pseudoscalar states at infinitely large energies were published in \cite{Chang1,Chang,Braaten1,Braaten}, while numerical evaluation confirming the analytic results allowed to observe conditions at which the fragmentation regime becomes justified \cite{Shevlyagin1,Shevlyagin}. So, the fragmentaion of heavy quark  to the vector state is more hard than one to the pseudoscalar state, that confirms the assumption on the more virtual heavy quark fragmenting to the vector meson.

Analogous perturbative calculations can be made for partonic processes of  gluon fusion and annihilation of light quark and anti-quark in hadronic processes that shows the regime of fragmentation becomes to dominate in comparison with a recombination to heavy quarkonium  at high transverse momenta $p_\perp > 6 \cdot m_{Q\bar Q}$, only \cite{Berezhnoy1,Berezhnoy}. 

Thus, there are the mentioned reasons for significant uncertainties in methods of extracting the soft fragmentation functions of heavy quarks due to the mixed mechanisms yielding the observed data on the heavy hadron distributions. Those are the accuracy of hard evolution spreading  exact initial data and recombination screening the fragmentation in regions, where cross sections are significant.

\section{Conclusion}
In this paper the substantiated analysis for the reciprocity relation between the heavy quark distribution in the heavy hadron and the heavy quark fragmentation function in the soft regime has been done. The qualitative features of such relations are in agreement with empirical data. However, the data require the separation of different mechanisms, that involves a significant uncertainties in any extraction of primary distributions for the fragmentation description.

\end{document}